\documentclass{iau}

\usepackage{amsmath}
\usepackage{graphicx}
\usepackage{multirow}
\begin{document}

\lefttitle{Kofman}
\righttitle{Spectroscopic signatures from the habitable zone}

\jnlPage{1}{7}
\jnlDoiYr{2026}
\doival{10.1017/xxxxx}

\aopheadtitle{Proceedings IAU Symposium}
\editors{J. Haqq-Misra \&  R. Kopparapu, eds.}

\title{Spectroscopic signatures from the habitable zone}

\author{Vincent Kofman}
\affiliation{Centre for Planetary Habitability, Department of Geoscience, University of Oslo, Oslo, Norway}

\begin{abstract}
This work describes the context and approach for the detection of spectroscopic signatures from planets in the habitable zone of nearby stars. By understanding the limitations of current observatories, future telescopes can be understood, and their ability to characterize the atmospheres of exoplanets estimated. An example calculation is given for the signal-to-noise analysis for a planet like the current Earth of oxygen as a biosignature, and (an enhanced abundance) of hydrogen iodine as a technosignature. In the optimistic estimate, Earth is easily detected, O$_2$ characterized in 20 hours, but signals from enhance HI are only visible after hundreds of hours, indicating the signals are too weak to realistically constrain. 
\end{abstract}


\maketitle

\section{Introduction}
The characterization of the surfaces and atmospheres of exoplanets is advancing rapidly as a research field. From broad spectral points with low signal-to-noise ratios using the \textit{Hubble Space Telescope}  \citep{deming_highlights_2020,de_wit_combined_2016}, observations using the \textit{James Webb Space Telescope} (JWST) now routinely explore the composition of atmospheres of exoplanets \citep{espinoza_highlights_2025}. Even well before the first discovery of exoplanets, speculation about finding life outside of the Earth has been rife throughout the scientific and science-fiction literature, but now we are progressing towards being able to actually constrain whether planets may or may not be habitable. Detailed discussions are available on what would constitute proof for habitability \citep{schwieterman_exoplanet_2018} and how to assess the ramifications of a potential detection \citep{green_call_2021,meadows_community_2022}. 

The focus of the work here is on the detectability of spectroscopic signatures from terrestrial planets in the habitable zone. In our current understanding, these are the best targets to find life as we are familiar with \citep{cawood_evolution_2025}. Within this paradigm, the detectability of spectroscopic signatures originating from biology and technology are explored. Technosignatures can be considered as 'advanced biosignatures', traces from civilization and potentially less ambiguous in their origins than biosignatures, and in some cases more detectable. 

Currently, observations of terrestrial planets in the habitable zone are limited to transiting planets around relatively cool stars. Around cooler, smaller stars, planets are closer in, increasing their odd of transit alignment and enhancing the signal of the transit with respect to the brightness of the star. The relatively cool stars are not considered to be favorable hosts for habitability however. The habitable zones around hotter stars are further from the stars, likely leading to more hospitable environments (e.g. see \citet{kopparapu_habitable_2018}). Efforts to understand $\eta_{Earth}$, the likelihood of habitable planets around Sun-like stars (3800-6800 Kelvin; K, G, F-type), vary considerable \citep{fernandes_are_2025}, ranging from 1 in 100 to close to up to 1. Observational constrains are totally lacking here however. Statistical analysis of the space-based Kepler survey, which currently provides our most complete census of terrestrial planets in the habitable zone, indicate that so far much less than 1 out of 10000 habitable zone planets would have been found \citep{bryson_probabilistic_2020}. The upcoming PLATO mission may be able to provide better constrains on $\eta_{Earth}$, being more sensitive than Kepler, and targeting the same stars for longer \citep{rauer_plato_2024}, but is still limited to transiting planets.

Constraining the current remote observing capabilities allows us to understand the possibilities of future observatories. It also helps us better grasp what is possible now, and perhaps treat extraordinary claims with the appropriate amount of skepticism. Here, the current capabilities of telescopes will be briefly explained and quantified in terms of the strength of the information content with respect to the brightness of the star. This enables predictions of the possibilities of potential future observatories such as the space-based Habitable Worlds Observatory and the Large Interferometer for Exoplanets, as well as the next generation of ground-based observatories. 

\begin{figure}
    \centering
    \includegraphics[width=.75\linewidth]{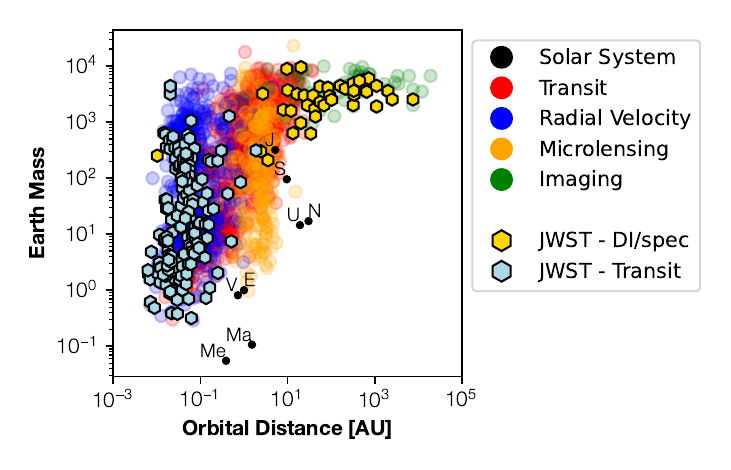}
    \caption{The mass versus orbital distance of currently discovered and characterized exoplanets. Solar system planets are indicated as well. The hexagonal icons, indicating spectroscopic characterization studied done using JWST, are adopted from \cite{espinoza_highlights_2025}. Direct imaging, currently limited to far out and large planets, will have to advance in sensitivity and contrast ratio to reach Earth-like planets.}
    \label{fig:exoplanets}
\end{figure}

In Figure \ref{fig:exoplanets}, the currently discovered exoplanets are shown as a function of their separation from their host star and the planets' mass. The different colors indicate the methods used to find these, which reveal where these techniques are sensitive to. Also shown are planets that have been studied with JWST, transit studies in light blue, direct imaging in yellow, adopted from \citet{espinoza_highlights_2025}. The points give an overview of the parameter space that is currently explorable, and by looking at a number of these observations, the sensitivity of current methods can be assessed. For the case of transit observations, the TRAPPIST-1 system has been of significant interest to the community. A series of observations analyzed with different data-reduction pipelines yielded errors in the 100 ppm range \citep{lim_atmospheric_2023} which is close to the maximum signal strength that would be expected for a modern Earth-like atmosphere \citep{fauchez_trappist-1_2022}. The issue is that the stellar variation remains a large factor of uncertainty in transit observations however, with differences between the several transit observations offset up to five times the noise level, indicating that the telescope noise currently may not be the limiting factor in the observations. Estimating from the simulated spectra, noise levels to meaningfully constrain the atmosphere of TRAPPIST-1 e should be in the 10s of ppm or $10^{-5}$ in units of contrast with respect to the host star, from which we can conclude that at least an order of magnitude improvement is required.

\section{Currently known exoplanets and facilities to detect and characterize these}

If we now turn to transit observations of Earth-size planets around the larger Sun-like stars, the contrast ratios need to be in the $10^{-7}$ in order to spectroscopically constrain their atmospheres (see for instance \citet{lustig-yaeger_earth_2023}). With larger separation and lower transit signals, studies aiming to characterize atmospheres of Earth-like planets around Sun-like stars are likely not feasible, particularly since the relative constant brightness of the Sun does not seem to be typical for G stars \citep{herbst_comparing_2025}.

\section{Future observatories and their potential to characterize Earth-like exoplanets}
\begin{figure}
    \centering
    \includegraphics[width=1\linewidth]{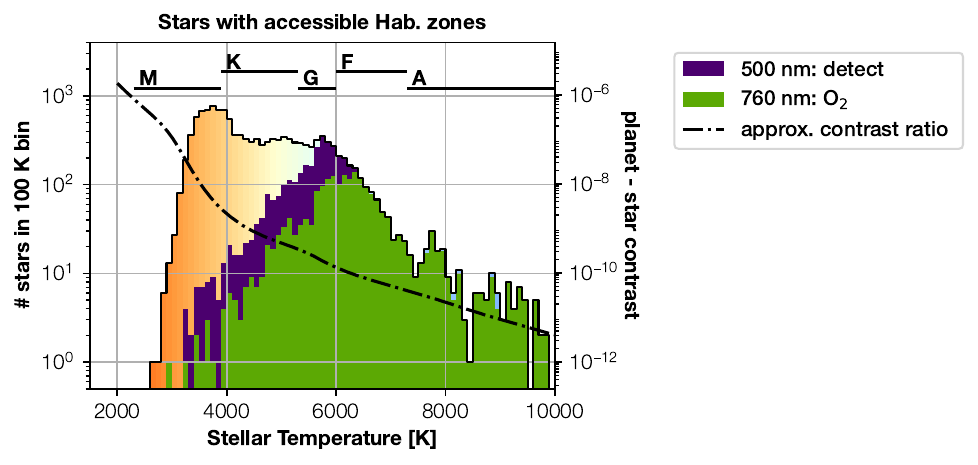}
    \caption{Sufficiently bright stars within with habitable zones accessible for a telescope 6.5 meter diameter telescope with sensitivity to planets at an inner working angle of 2$\frac{\lambda}{D}$. The approximate planet-star contrast of a Earth-size planet in the habitable zone is shown. The planet-star flux scales as $A\cdot[\frac{P_r}{a}]^2$,  with $A$ the planets' reflectivity at quadrature, $P_r$ the planet radius, and $a$ the orbital separation.} 
    \label{fig:stars}
\end{figure}

Since signals from transiting Earth-like planets around Sun-like stars seem to be too weak compared to stellar variability, this brings us to the possibility for the study of Earth-like planets in either reflected or emitted light. In both cases, the stellar light will have to be removed from the observations. Regardless of the methods used (interferometric, coronagraph, or starshade \textit{e.g.} see \citet{serabyn_stellar_2024}), there are commonalities that can be explored between the techniques. The separation on the sky fundamentally limits the separation of light from the star and planet, with the wavelength of light $\lambda$ and the diameter of the telescope $D$ determining the size of the diffraction pattern ($\sim$ 1.22$\frac{\lambda}{D}$;  the Airy ring). An application of this limit is demonstrated in Figure \ref{fig:stars}. Shown are stars that are bright enough to be observed, in this case assumed as brighter than 12th magnitude, and within 50 parsec of the Sun \citep{tuchow_hpic_2024}. Taking into consideration the distance for each of the stars in the sample, we can constrain whether we can probe the habitable zone of this star considering the physical $\frac{\lambda}{D}$ limitation. Two wavelengths are considered, corresponding to the initial detection band at 500 nm and to absorption from O$_2$ at 760 nm. One iteration of the future Habitable Worlds Observatory is considered, with an inner working angle of 2$\frac{\lambda}{D}$ and a mirror size of 6.5 meter. For stars hotter than our sun, the habitable zone can be studied in almost all cases, but the detectability of O$_2$ is already more limited. Studying stars much colder than 5000 K becomes incidental, as the inner working angle precludes this. Also shown is the approximate contrast ratio between the planet and the star, around 1.8$\cdot$10$^{-10}$ for Earth-Sun, but increasing for colder stars, which may offer an opportunity to characterize planets around K-stars.

In terms of current ground-based telescopes, the largest telescopes are 8 meters (Very Large Telescope), but instruments are not able to reach the contrast levels required. The next generation of ground-based observatories, the Giant Magellan Telescope and the European Large Telescope, with their larger mirror sizes of 22 and 38 meters respectively, will enable probing of the habitable zones of almost all stars in the sample. Pathways to obtain the contrast ratios over these distances are being developed \citep{kenworthy_high-contrast_2025}. Somewhat different from space-based telescopes, instruments from ground-based observatories can undergo many iterations after the telescopes becomes operational. As high-resolution spectroscopy is only possible from the ground, these new observatories hold immense potential for finding signatures from the habitable zone. 

\section{The detectability of spectral features}
In order for spectroscopic signatures from a molecule, atom, or surface to be \textit{detectable} and \textit{identifiable}, the follow criteria should be considered. Controversies arise when either of the two criteria are not appropriately met.

1) The feature needs to be significantly offset from the continuum. What significant means can be expressed using signal-to-noise (SNR) analysis by running a forward model with and without the signal and assessing the SNR of the feature. Examples will be given below. \textit{Detections} are typically quantified as a SNR larger than five. Alternatively, retrieval studies can be performed for the exploration of potential overlap of features \citep{line_systematic_2013}. It should be noted that the basic SNR analysis provides a solid constraint on the upper limit of the detectability if no overlaps are assumed. 

2) Once a spectral feature of significant strength is detected, it needs to be sufficiently uniquely \textit{identifiable} to constrain its origin. Many molecules share functional groups (e.g. CH or CO), which manifest in same range in the infrared. Higher resolution studies, or validations of features over a broader spectral range can help here.
\subsection{Examples of spectroscopic signatures}
\begin{figure}
    \centering
    \includegraphics[width=1\linewidth]{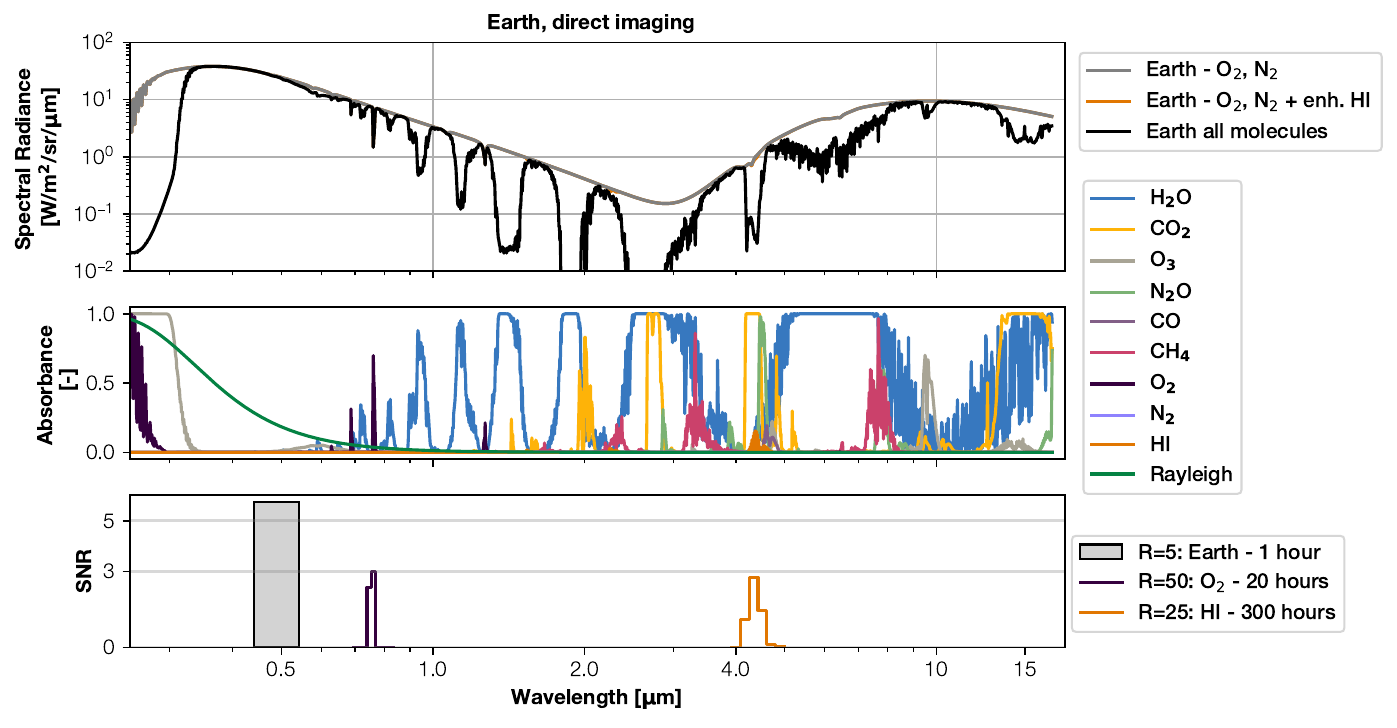}
    \caption{Top panel: reflected and emitted light spectrum of a cloudless Earth, based on \citet{haqq-misra_challenge_2024}. Middle interactions of difference molecular processes as a function of wavelength. Bottom panel, SNR of different signals at the indicated resolving power.}
    \label{fig:spectrum}
\end{figure}

To understand the context of the spectral signatures, in Figure \ref{fig:spectrum} the reflected and emitted light spectrum of the Earth is shown. The figure is based on  work in \citet{haqq-misra_challenge_2024}, where a radioactive isotope of HI was considered as a technosignature. Reflected light dominates the spectrum to about 2.7 $\mu m$, after which the emission of the Earth and its atmosphere is brighter than reflected Sunlight. In the top panel, three curves are shown. First, Earth with only O$_2$ and N$_2$ in gray, then in black the Earth including its atmosphere and all relevant processes, except clouds. The orange trace shows the signals from HI in a N$_2$ and O$_2$ only atmosphere, as other species overlap (as of such it is not a detectable species, even at the enhanced abundance investigated). The traces are in units of spectral radiance, W m$^{-2}$ sr$^{-1}$ $\mu m^{-1}$. The middle panel shows the interaction of different species and processes with the light, helping to identify the absorption features in the top panel. In the infrared, we see that light mostly does not propagate through the atmosphere, so the bulk of the emission comes from the atmosphere itself. The bottom panel shows the SNR of three different detections using a Habitable Worlds Observatory analogue \citep{kofman_pale_2024}, and a telescope similar to the Large Interferometer for Exoplanets \citep{dannert_large_2022}.

Looking at the reflected part of spectrum, we can see why one would target signals from O$_2$, H$_2$O, and possibly O$_3$. O$_2$ is narrow, very unique, and in a relatively bright, uncrowded area. H$_2$O is present throughout the spectrum, but the spectrum is already 10 times fainter above 1 $\mu m$. O$_3$ is a strong absorber, even at its parts per billion abundance, but the absence of light is not a uniquely identifiable as the feature is very broad. Finally, even though the Sun emission peaks around 550 nm, the 300-500 nm range is brighter due to Rayleigh scattering, indicative of an atmosphere. In the infrared, H$_2$O, O$_3$, and CO$_2$ are the clearest signals. Clouds affect the spectra strongly, affecting the detectability of signals in reflected light \cite{kofman_simulating_2026}. 

\subsection{Estimating the signal-to-noise ratio of features based on the photon flux}
The noise level of the observation can be estimated from the expected photon flux. Often, a noise level that is constant with wavelength is assumed, but this is unrealistic as the planet is much less bright at longer wavelengths. Here, we will calculate the photon flux dependent noise level for three signals, and estimate the required observational time required to confidently detect these. First, the detection of the planet at 500 nm will be investigated, then the presence of O$_2$ constrained, and finally the potential HI signal is studied.

If we were to observe the solar system from a distance of 10 parsec, the absolute bolometric magnitude Sun is +4.83, which can be converted in photons per second per micron using the following conversions. AB magnitude is defined with respect to spectral irradiance (or flux density in the astronomical literature) $f_{\nu}$ in ergs cm$^{-2}$ s$^{-1}$ Hz$^{-1}$ \cite{oke_secondary_1983}:
\begin{equation}
    AB = -2.5 \log f_{\nu} - 48.60
    \label{eq:placeholder_label}
\end{equation}
This can be inverted to obtain the spectral irradiance, now in units of W m$^{-2}$ Hz$^{-1}$:
\begin{equation}
f_{\nu} = 10^{-3} \cdot  10^{-\frac{AB + 48.60}{2.5}}   = 10^{-\frac{AB + 56.10}{2.5}} 
\end{equation}
where the factor $10^{-3}$ converts the units from ergs cm$^{-2}$ s$^{-1}$ Hz$^{-1}$ to W m$^{-2}$ Hz$^{-1}$. This yields 4.246 $\cdot10^{-25}$ for our example. Now to convert the units to the more familiar W m$^{-2}$ $\mu$m we use the following equation \cite{viana_near_2009}:

\begin{equation}
    F_{\lambda} = F_{\nu} \cdot \frac{c}{\lambda^{2}} = 4.246 \cdot10^{-25}\frac{c}{0.5556^2} 10\cdot10^{6}= 4.124\cdot10^{-10} \space W m^{-2}\mu m
    \label{eq:placeholder_label}
\end{equation}

Where $\lambda$=0.5556 $\mu m$. The spectral irradiance can be converted to photons s$^{-1}$ by dividing by the photon energy ($\frac{hc}{\lambda}$).
Thus,
\begin{equation}
4.124\cdot10^{-10} \cdot \frac{\lambda}{hc} = 1.142\cdot10^{9} \space ph \space s^{-1} \space m^{-2}\mu m
\end{equation}
Finally, considering the Earth-to-Sun contrast ratio is 1.8 $\times 10^{-10}$ can estimate the photon received a telescope with mirror size $r$ and band pass width $\Delta\lambda$.
\begin{equation}
\frac{1}{2}\pi r^{2} \cdot\Delta\lambda \cdot 1.142\cdot10^{9} \cdot 1.8\cdot10^{-10}= 0.12\space ph \space s^{-1} \space \mu m
\end{equation}
In order to detect the planet, the photons can essentially be binned to the width of the coronagraph bandpass,  assuming 20\%, this is approximately 0.1 $\mu m$. The signal to noise ratio can be obtained by the following equation, which also considers the quantum efficiency $QE$, the detector dark $N_d$ and readout noise $N_r$

\begin{equation}
    \mathrm{SNR} = \frac{I \cdot QE \cdot t}{\sqrt{I \cdot QE \cdot t + N_d \cdot t + N_r^2}},
    \label{eq:placeholder_label}
\end{equation}
with $I$ our photon flux. Substituting $QE=0.27$, $N_d = 3\cdot10^{-5}$, $N_r=0.008$ \citep{kofman_pale_2024}, $t=3600$ and $I = 0.12 \cdot 0.1$ yields us an SNR of 10.8. This is optimistic, as it does not consider speckle noise, (exo)zodiacal noise, or any other non-photon noise observing challenges, but it provides an upper limit to the spectroscopic signal. 

To assess the SNR of spectral signals \textit{within} the planet's spectrum, we need to assess the difference between the spectrum with and without the signal, and express this with respect to the noise. From \citet{kofman_pale_2024}, where $m_{s,k}$ is the spectrum with the signal, $m_{ref,k}$ is the reference signal, $\sigma_k$ is the noise, all expressed at wavelength $k$: 
\begin{equation}
    \label{eq:placeholder_label}
    \mathrm{SNR}_{s-w} = \sqrt{\sum_{k} \left( \frac{m_{s,k} - m_{w,k}}{\sigma_{k}} \right)^{2}}
\end{equation}
In the case of the O$_2$ and the HI observations, we need to choose a suitable resolving power that enables the feature to be distinguishable. Note here the difference between the resolving power $R$ and the resolution $\Delta\lambda$. The two are related as following: $R=\frac{\lambda}{\Delta\lambda}$. Higher resolving powers lead better identifiable signals but a proportional decrease in the photon count. 

We can see in figure \ref{fig:spectrum} that the enhanced signals from HI, even optimistically assuming no overlap from CO$_2$ and H$_2$O, are too weak to be detected.

\section{Conclusion and outlook}
The examples above outlined the intricacies of studying spectral signals from the habitable zone. With the methods outlined above, estimations can be made for all kinds of signals from atmosphere of planets. If other signals want to be investigated (\textit{i.e.}, structures beyond the planet), the calculations provide a framework in which to place other simulations by comparing to the brightness of the planet. The addition of clouds enhances the brightness of spectral signatures further, see \citet{kofman_simulating_2026}. 

Characterizing the atmospheres of small planets that are potentially habitable is likely to become possible with our lifetimes. Preparing for these developments is critical as it will help shape the expectations of the community and possibly guard against claimed detections that may not meet the criteria for rigorous detections of signatures. 

\bibliography{references}{}
\bibliographystyle{aasjournal}

\end{document}